\documentclass{jfm}

\usepackage{graphicx}
\usepackage{epsfig}
\usepackage{natbib}
\usepackage{amssymb}
\usepackage{upmath}
\usepackage{amsbsy}
\usepackage{epstopdf}

\usepackage{amsmath}
\usepackage{url}
\usepackage{graphicx}
\usepackage{verbatim}
\usepackage{subfigure}
\usepackage{subfig}

\newcommand{\be}{\begin{equation}}
\newcommand{\ee}{\end{equation}}
\newcommand\St{\mbox{\textit{St}}}  
\newcommand\Ca{\mbox{\textit{Ca}}}  
\newcommand\We{\mbox{\textit{We}}}  

\newcommand\mum{\nobreak\mbox{$\;\umu$m}}

\newcommand{\dx}{\mathrm{d}}

\title[Universal mechanism for air entrainment during liquid impact]{Universal mechanism for air entrainment during liquid impact}

\author[M.H.W. Hendrix, W. Bouwhuis et al.]{Maurice H.W. Hendrix$^{1,2}$, Wilco Bouwhuis$^{1}$, Devaraj van der Meer$^{1}$, Detlef Lohse$^{1}$, and Jacco H. Snoeijer$^{1,3}$ }

\affiliation{$^1$ Physics of Fluids Group, Faculty of Science and Technology, University of Twente, 7500 AE Enschede, The Netherlands,\\$^2$ Laboratory for Aero and Hydrodynamics, Delft University of Technology, Leeghwaterstraat 21, NL-2628 CA Delft, The Netherlands,\\$^3$ Mesoscopic Transport Phenomena, Eindhoven University of Technology, Den Dolech 2, 5612 AZ Eindhoven, The Netherlands}

\pubyear{2012}
\volume{?}
\pagerange{??}
\begin{document}

\maketitle

\begin{abstract}
When a mm-sized liquid drop approaches a deep liquid pool, both the interface of the drop and the pool deform before the drop touches the pool. The build up of air pressure prior to coalescence is responsible for this deformation. Due to this deformation, air can be entrained at the bottom of the drop during the impact. We quantify the amount of entrained air numerically, using the Boundary Integral Method (BIM) for potential flow for the drop and the pool, coupled to viscous lubrication theory for the air film that has to be squeezed out during impact. We compare our results to various experimental data and find excellent agreement for the amount of air that is entrapped during impact onto a pool. Next, the impact of a rigid sphere onto a pool is numerically investigated and the air that is entrapped in this case also matches with available experimental data. In both cases of drop and sphere impact onto a pool the numerical air bubble volume $V_b$ is found to be in agreement with the theoretical scaling $V_b/V_{drop/sphere}\sim \St^{-4/3}$, where $\St$ is the Stokes number. This is the same scaling that has been found for drop impact onto a solid surface in previous research. This implies a universal mechanism for air entrainment for these different impact scenarios, which has been suggested in recent experimental work, but is now further elucidated with numerical results.
\end{abstract}

\section{Introduction}\label{sec1}

\begin{figure}
	  \centerline{\includegraphics[width=.9\textwidth]{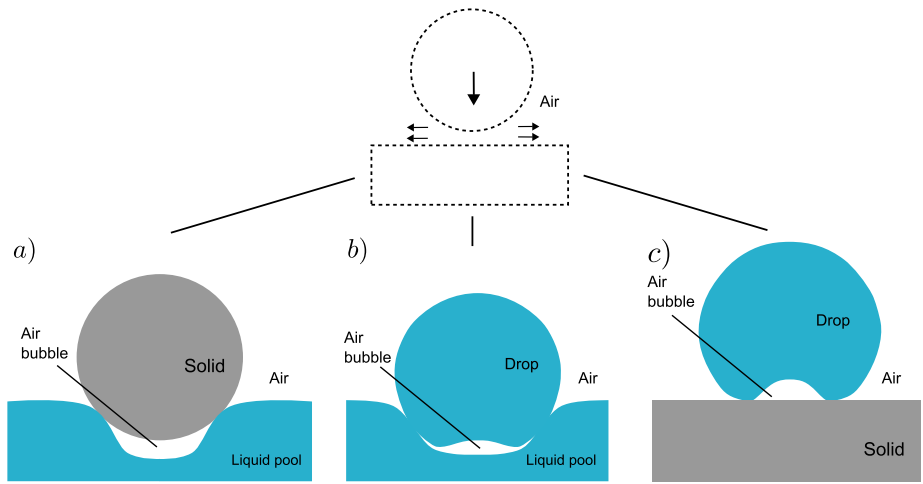}}
		\caption{Air bubble entrapment for different impact scenarios. Bubbles and deformations are not drawn to scale. (\textit{a})  Rigid sphere impact onto a pool. The pool deforms due to an increase in air pressure right under the sphere before it touches the pool, which results in an entrapped air bubble. (\textit{b}) Drop impact onto a pool. Not only the pool, but also the drop consists of a deformable interface. As a result, the increased air pressure deforms both the pool and the drop and an air bubble is entrapped. (\textit{c}) Drop impact onto a solid. Also here, a local increase in air pressure deforms the drop before it touches the solid and results in an entrapped air bubble.}
		\label{fig:air}
	\end{figure}

The impact of a drop or a solid sphere onto a liquid pool can encompass various types of air entrainment. One possibility is that air is entrained at the top of the impacting object when the crater that is created during impact collapses, see for example~\cite{Oguz_1990,Pumphrey_1990,Wang_2013,Chen_2014}. Another type of air entrainment may occur at the bottom of the impacting object: the thin air film that is squeezed out at the impact zone is accompanied by a pressure increase that deforms the interface of the liquid before the impacting object touches the pool, which may result in air entrapment~\citep{Marston_2011,Tran_2013}. The early stages of deformations can be described analytically~\citep{Bouwhuis_2015}. In case the impacting object is a drop, instead of a single entrapped bubble, also a collection of microscopic bubbles may be entrapped which can create intriguing morphologies~\citep{Thoroddsen_2012}. This is also referred to as Mesler entrainment~\citep{Esma_1986,Pumphrey_1990}. The same mechanism that is responsible for bubble entrapment at the bottom of an impacting object on a pool holds for of air entrapment at the bottom of an impacting drop onto a solid~\citep{Dam_2004,Mani_2010,Hicks_2010,Bouwhuis_2012}.  In fact, the initial geometry of the problems is identical, see figure~\ref{fig:air} in which the different impact scenarios and air entrapment have been depicted. We also refer to figure 5 of~\cite{Tran_2013}, who first worked out this analogy. 

Previously, air bubble entrapment for drop impact onto a solid surface has been quantified experimentally, theoretically, and numerically~\citep{Mandre_2009,Mani_2010, Hicks_2010,Hicks_2011,Bouwhuis_2012}. If the effect of surface tension can be neglected, the following scaling for the entrapped air bubble volume was found: 

\be
V_b/V_{drop}\sim St^{-4/3}.
\label{vb2}
\ee 

\noindent Here $V_b/V_{drop}$ is the air bubble volume normalized by the drop volume and $\St$ is the Stokes number which is defined as $St\equiv\rho_l R U/\eta_g$, where $\rho_l$ is the liquid density, $R$ the droplet radius, $U$ its impact velocity, and $\eta_g$ is the viscosity of the surrounding gas, in this case air. The Stokes number represents the competing effect of the viscous force of the draining air film and the inertial force of the liquid which ultimately determines the air bubble volume. The same scaling was found experimentally for impact of a sphere onto a pool~\citep{Marston_2011}, and a drop onto a pool~\citep{Tran_2013}. 

In this paper we try to capture the mechanism of air entrapment during impact onto a pool numerically. We will employ a boundary integral method (BIM) for potential flow describing the liquid phase coupled to viscous lubrication theory for the draining microscopic air film. The advantage of using a boundary integral method becomes evident when the interface of the impacting object comes close to the pool and one has to resolve the microscopic air layer together with the macroscopic liquid scale. This difference in length scale can be a thousandfold for the case of a millimeter sized drop impacting onto a pool squeezing out an air film with a typical thickness of micrometer. In fact, the difference in length scale in the final stages of impact diverges to infinity as the drop is about to coalesce with the pool. Using a boundary integral method guarantees excellent interface representation, since all variables such as liquid velocity and pressure are defined at the interface. At the same time, the computational cost is modest, since the boundary integral method allows the potential problem to be solved only at the boundaries of the liquid domain: quantities in interior points can be calculated optionally as a function of the solution at the boundary. To achieve the same accurate interface representation and solving the full Navier-Stokes equations, using for example a volume-of-fluid method (see for example~\cite{Thoraval_2012,Guo_2014}), would require a much larger computational cost. 

In section~\ref{sec:numerical} we explain the theoretical framework together with the numerical method. In section~\ref{sec:results} we will present the results of numerical simulation: we will identify details of the pressure development in the air film and deformation of the interfaces at the impact zone. The results of the numerical model will be compared with available results regarding the entrapped bubble volume from multiple experimental works and will be compared with the scaling law equation (\ref{vb2}). We conclude with section~\ref{sec:conclusion} in which also suggestions for further research are discussed.  



\section{Theory} \label{sec:numerical}
\subsection{Dimensional analysis and numerical method}
The Reynolds number of the liquid drops we model, which is defined as $Re_{l}\equiv \rho_{l}RU/\eta_l$, is assumed to be large, $Re_{l}\gg 1$. Here $\rho_l$ and $\eta_l$ are respectively the density and the dynamic viscosity of the liquid, $U$ is the impact velocity and $R$ is the radius of drop. The flow can thus be regarded irrotational, that is $\nabla \times {\bf u}={\bf 0}$. Under the additional constraint of incompressible flow inside the drop this allows the liquid dynamics to be modeled with a harmonic function $\phi$, to which the velocity field ${\bf u}$ is related through:

\be
{\bf u}=\nabla \phi
\label{grad}
\ee
 
\noindent The fact that the velocity potential $\phi$ obeys the Laplace equation $\nabla^2\phi=0 $ is used to efficiently solve the potential problem, and thus the dynamics of the liquid, using the Boundary Integral Method (BIM). While the Reynolds number of the drop is large, the Reynolds number of the thin gaseous air layer $Re_{g}\equiv \rho_{g}H_{d}U/\eta_g$ is typically small, $Re_{g}\ll 1$. Here $ \rho_{g}$ is gas density and $H_d$ is the air film thickness in the center of the film which is referred to as the dimple height. The length scale characterizing the air layer in the lateral extension of the air film is denoted by $L$, see figure~\ref{fig:tn}a. As shown in ~\cite{Bouwhuis_2012}, $H_d\ll L$ which in combination with the low Reynolds number of the gas allows the film to be described with viscous lubrication theory, see for example~\citet{Leal1992345}. The dimensionless group reflecting the presence of air is the Stokes number $\St\equiv \rho_{l}RU/\eta_g$ which compares the viscous force of the air layer to the inertial force in the drop. This number is relevant for describing dimple formation, since, for high enough impact velocity $U$, this process is determined by two competing forces: the force of the viscous air layer trying to deform the drop in the center and the opposing inertial force of the drop, which must be slowed down locally in order to form a dimple. Additional dimensionless numbers incorporating surface tension $\gamma$ are the Weber number $\We$ and the capillary number $\Ca$ based on the gas properties. Summarizing, we thus have the following dimensionless parameters:

\begin{equation}\label{dimensionless}
  Re_{l}\equiv \frac{\rho_{l}RU}{\eta_l}  \hspace{1cm} Re_{g}\equiv \frac{\rho_{g}H_{d}U}{\eta_g}  \hspace{1cm} St\equiv \frac{\rho_{l}RU}{\eta_g} \hspace{1cm} We\equiv \frac{\rho_l R U^2}{\gamma}  \hspace{1cm}  Ca\equiv \frac{We}{St} 
\end{equation}

\noindent The impact of a liquid drop onto a pool of the same liquid and the impact of a rigid sphere onto a liquid pool can be described with the same dimensionless numbers. As the initial geometry of the problems is identical, the difference lies in the deformability of the object, which is zero in case of the solid. The two effective control parameters that we will use here in our theoretical framework are $\St$ and $\We$.

In figure~\ref{fig:tn}a an illustration of the impact of a drop onto a pool, together with the used method is shown. As is clear from this figure, the coupling between the dynamics of the air layer and the dynamics of the liquid is essential since the two liquid domains feel each other through the pressure build up in the viscous air layer. The lubrication pressure $P_g$ acts on the liquid surface and appears in the unsteady Bernoulli equation which serves as a boundary condition in the BIM which is applied at the liquid surface, see also~\citet{Bouwhuis_2012}. As we have two liquid domains, two separate BI equations are solved. We take the width of the pool large enough to approach the dynamics of an infinite liquid pool. In this case a width of 4.5 times the drop radius was found to be sufficient. We focus on quantifying the amount of entrapped air by integrating the enclosed air pocket up to the moment the air layer reaches a physical minimum thickness of $0.4\mum$. At this point the volume of the enclosed air has converged and a subsequent rupture of the air film will prevent further drainage which results in an entrapped air bubble~\citep{Bouwhuis_2012}. As we focus on the dynamics just prior to rupture we can make use of an axisymmetric framework. We restrict ourselves to the inertial regime for which experimental results~\citep{Tran_2013} are available for a direct comparison. Since the air layer continually deforms and translates during the impact, lubrication equations have been developed in a moving coordinate system which is aligned with the interface of the drop. These equations will be derived in the next section. 

\subsection{Lubrication in moving and tilted coordinate system}

\begin{figure}
	\centering
		 \includegraphics[width=0.9\textwidth]{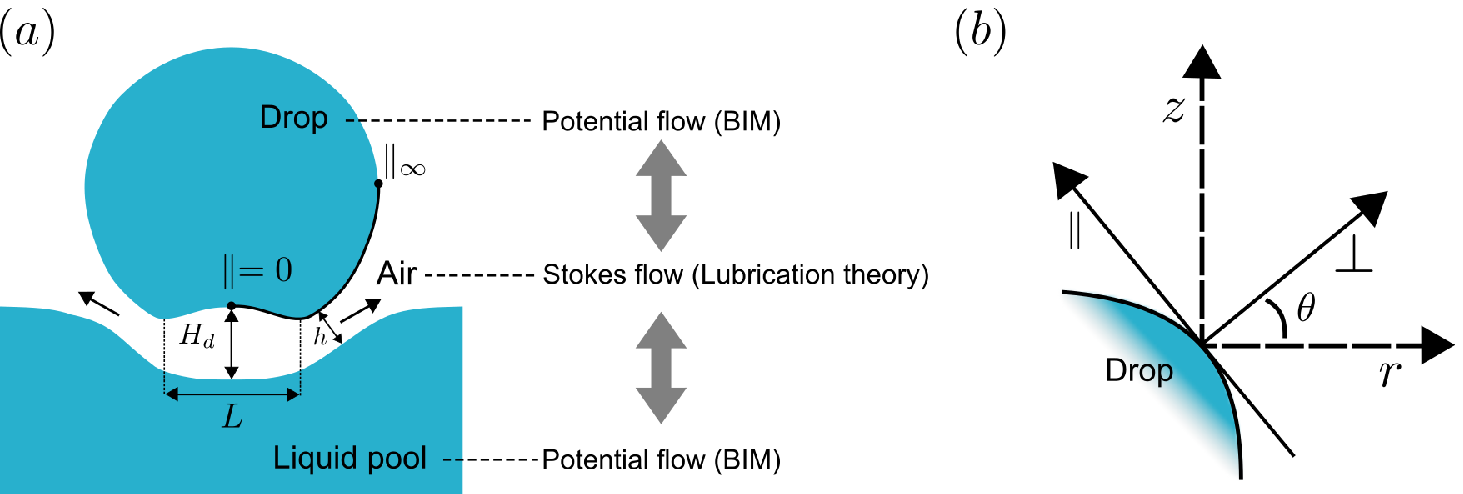}
		\caption{(\textit{a}) Schematic of drop impact onto a a pool. The used methods are indicated in the figure: both the liquid domains are modelled with potential flow, while the air layer is described with Stokes flow. The gray arrows indicate that the flow of the air film is coupled to the dynamics of the liquid domains and vice versa. (\textit{b}) Definition of the ($\parallel$,$\perp$)-coordinate system where $\parallel$ is aligned along the drop curve and $\perp$ is the unit normal with respect to the drop.}
		\label{fig:tn}
	\end{figure}

In this section we develop an expression for the pressure $P_g$ in the air film based on lubrication theory in a moving ($\parallel$,$\perp$)-coordinate system which is aligned along the drop surface, see the sketch in figure~\ref{fig:tn}. The reason for doing this (rather than just using the standard ($r$,$z$)-coordinate system) is that especially for the drop onto pool impact, the moving ($\parallel$,$\perp$)-coordinate system is not necessarily oriented as the ($r$,$z$)-coordinate system and therefore only the first guarantees an accurate description of the draining air film. The drop surface is taken as a reference, and the curvilinear coordinate $\parallel$ is defined along the drop, starting at the axis of symmetry.  At some large radial coordinate $\parallel_{\infty}$ we assume atmospheric pressure. The coordinate perpendicular to $\parallel$ is defined to be $\perp$. The gap height $h(r,t)$ is defined as the length of the perpendicular line from the drop projected onto the liquid pool. The two surfaces in the impact zone are assumed to be nearly parallel ($\left |{\partial_{\parallel}}h  \right |\ll 1$), so we can apply lubrication theory. 

It can be shown (see appendix~\ref{sec:app}) that the continuity equation in this new ($\parallel$,$\perp$)-coordinate system reads:

\be
\frac{u_{r}}{r}+\partial_{\parallel}u_{\parallel}+\partial_{\perp}u_{\perp}=0.
\label{continuity_tn}
\ee

\noindent At the interface of the liquid pool ($\perp=h$) we know that the fluid particles have to move with the interface. This is mathematically described with the kinematic boundary condition:

\begin{equation}\label{kinematic}
\partial_{t}h + \left (  u_{\parallel}\left. \partial_{\perp}h \right )\right|_{\perp=h}=u_{\perp\mid \perp=h}-u_{\perp\mid \perp=0}.
\end{equation}

\noindent Here $\partial_{t}h$ is the time derivative of $h$. We now integrate equation (\ref{continuity_tn}) along the gap height $h$ and obtain:

\begin{equation}\label{continuity_int}
\int_{0}^{h}\frac{u_{r}}{r}\,\dx\perp+\int_{0}^{h}\partial_{\parallel}u_{\parallel} \,\dx\perp=-\int_{0}^{h}\partial_{\perp}u_{\perp}\,\dx\perp=u_{\perp\mid \perp=0}-u_{\perp\mid \perp=h}.
\end{equation}

\noindent Using Leibniz integral rule for the second integral on the left hand side we find:

\begin{equation}\label{continuity_int_leib}
\int_{0}^{h}\frac{u_{r}}{r}\,\dx\perp+\partial_{\parallel} \int_{0}^{h}u_{\parallel}\,\dx\perp - \left (  u_{\parallel}\left. \partial_{\parallel}h\right )\right|_{\perp=h} = u_{\perp\mid \perp=0}-u_{\perp\mid \perp=h}.
\end{equation}

\noindent We now use the kinematic boundary condition formulated in equation (\ref{kinematic}) for the third term on the LHS to obtain:

\begin{equation}\label{continuity_int_leib}
\int_{0}^{h}\frac{u_{r}}{r}\,\dx\perp+\partial_{\parallel}\int_{0}^{h}u_{\parallel}\,\dx\perp + \partial_{t}h - u_{\perp\mid \perp=h} + u_{\perp\mid \perp=0} = u_{\perp\mid \perp=0}-u_{\perp\mid \perp=h}.
\end{equation}

\noindent Canceling the terms $u_{\perp\mid \perp=h}$ and $u_{\perp\mid \perp=0}$ on both sides gives:

\begin{equation}\label{continuity_int_leib}
\int_{0}^{h}\frac{u_{r}}{r}\,\dx\perp+\partial_{\parallel}\int_{0}^{h}u_{\parallel}\,\dx\perp + \partial_{t} h = 0.
\end{equation}

\noindent We still have to describe $u_{r}$ within the new ($\parallel$,$\perp$)-coordinate system. Therefore we substitute $u_{r}=u_{\perp}\cos \theta-u_{\parallel}\sin \theta$ in the equation above to get:

\begin{equation}\label{mass34}
\int_{0}^{h}\frac{1}{r}u_{\perp}\cos \theta\,\dx\perp - \int_{0}^{h}\frac{1}{r}u_{\parallel}\sin \theta\,\dx\perp + \partial_{\parallel}\int_{0}^{h}u_{\parallel}\,\dx\perp + \partial_{t} h= 0.
\end{equation}

\noindent We assume that the main flow of the air that is squeezed out from the gap is along the $\parallel$ coordinate, which implies  that $u_{\perp}$ is relatively small, so we neglect the first term. The second term is an integral with respect to $\perp$ containing the variable $r$. This radial coordinate $r$ across $h$ is a function of $\perp$: $r=\perp\cos\theta + c(\parallel)$. Here $c(\parallel)$ is the value of $r$ at the drop surface ($\perp=0$) for some coordinate $\parallel$. We thus substitute this expression for $r$ into equation (\ref{mass34}) and neglect the first term to find:

\begin{equation}\label{mass35}
 - \int_{0}^{h}\frac{\sin \theta }{\perp\cos\theta+c(\parallel)}u_{\parallel}\,\dx\perp + \partial_{\parallel}\int_{0}^{h}u_{\parallel}\,\dx\perp+\partial_{t} h= 0.
\end{equation}

\subsubsection{Flow profile within the air film}

\noindent As has been previously described the Reynolds number of the thin air film is small, $Re_{g}\ll 1$, and the geometry of the problem, $H_d\ll L$, allows us to use lubrication theory. In the ($\parallel$,$\perp$)-coordinate system, the Stokes equations can then be reduced to:

\begin{equation}\label{lub12}
\partial_{\parallel} P_g=\eta_g \partial_{\perp}^2u_{\parallel}. 
\end{equation}

\noindent We can integrate equation equation (\ref{lub12}) twice with respect to $\perp$, employing a no slip boundary condition at the drop surface ($u_{\parallel}=U_{d}$ at $\perp=0$) as well as at the surface of the pool ($u_{\parallel}=U_{p}$ at $\perp=h$):

\begin{equation}\label{pres}
u_{\parallel}= \left( (U_{p}-U_{d})\frac{\perp}{h}+U_{d} \right )+\frac{1}{2\eta_g}\partial_{\parallel}P_g(\perp^2-\perp h).
\end{equation}

\noindent The first term of equation (\ref{pres}) can be associated with Couette flow, caused by the movement of the interfaces. The second term can be associated with Poiseuille flow, which is driven by the radial pressure gradient, see also~\cite{Klaseboer_2000}. Substituting this expression for $u_{\parallel}$ in our equation for mass conservation, equation (\ref{mass35}), we get:

\begin{align}\label{mass335}
\notag - \int_{0}^{h}\frac{\sin \theta }{\perp\cos\theta+c}\left [\left( (U_{p}-U_{d})\frac{\perp}{h}+U_{d} \right )+\frac{1}{2\eta_g}\partial_{\parallel}(\perp^2-\perp h)  \right ]\,\dx\perp
\\
 + \partial_{\parallel}\int_{0}^{h}\left [\left( (U_{p}-U_{d})\frac{\perp}{h}+U_{d} \right )+\frac{1}{2\eta_g}\partial_{\parallel}P_g(\perp^2-\perp h) \right ]\,\dx\perp+\partial_{t} h= 0.
\end{align}

\noindent In the first integral we deal with a prefactor $\sin \theta / (\perp\cos\theta+c)$. When taking the geometry of the problem into account we note that $\perp\cos\theta \ll c$. We can thus write $\sin \theta / (\perp\cos\theta+c) \approx \sin \theta / c$. Performing the integrals of equation (\ref{mass335}) under this assumption yields: 

\begin{align}\label{mass445}
-\frac{\sin \theta }{c} \left (\frac{h}{2}\left ( U_{p} + U_{d} \right )-\frac{h^3}{12\eta_g}\partial_{\parallel}P_g    \right )  + \partial_{\parallel}\left (\frac{h}{2}\left ( U_{p} + U_{d} \right )-\frac{h^3}{12\eta_g}\partial_{\parallel}P_g  \right ) + \partial_{t} h=0.
\end{align}

\noindent If we define $G(\parallel)\equiv\left (\frac{h}{2}\left ( U_{p} + U_{d} \right )-\frac{h^3}{12\eta_g}\partial_{\parallel}P_g   \right )$ we can transform the above equation into a first order inhomogeneous linear ODE for $G(\parallel)$:

\begin{equation}\label{diff1}
\dot G(\parallel)-a(\parallel)G(\parallel)=f(\parallel).
\end{equation}

\noindent Here $a(\parallel)$ and $f(\parallel)$ are known functions of $\parallel$:

\begin{equation}\label{a}
a(\parallel)=\frac{\sin \theta }{c(\parallel)} 
\end{equation}

\begin{equation}\label{f}
f(\parallel)=-\partial_{t}h 
\end{equation}

\subsubsection{Solving the first order inhomogeneous ODE for $G(\parallel)$}

\noindent Equation (\ref{diff1}) can be solved with help of an integrating factor $I$ defined as $I(\parallel)\equiv\mathrm e^{-\int a(\parallel)d\parallel}$. Using the boundary condition $G(\parallel)=0$ for $\parallel=0$, because we have zero pressure gradient in the center of symmetry, and also zero tangential velocities, we can multiply equation (\ref{diff1}) with $I (\parallel)$ and solve for $G(\parallel)$:

\begin{equation}\label{ode3}
G(\parallel)=\frac{1}{I(\parallel)} \left( \int_{0}^\parallel I(\tilde \parallel)f(\tilde \parallel)\,\dx \tilde \parallel \right)
\end{equation}

\noindent with $I(\parallel)=\mathrm e^{-\int_{0}^{\parallel} a(\tilde \parallel)\,\dx \tilde \parallel}$.

%
%
%
%
%

\noindent We can now substitute $G(\parallel)\equiv\left (\frac{h}{2}\left ( U_{p} + U_{d} \right )-\frac{h^3}{12\eta_g}\partial_{\parallel}P_g   \right )$ back into equation (\ref{ode3}) to obtain an equation for $\partial_{\parallel}P_g   $:

\begin{equation}\label{ode4}
 \partial_{\parallel}P_g =-\frac{12\eta_g}{h^3}\left( \frac{1}{I(\parallel)} \left( \int_{0}^{\parallel} I(\tilde \parallel)f(\tilde \parallel)\,\dx\tilde \parallel  \right)-\frac{h}{2}\left( U_{p}+U_{d} \right) \right)
\end{equation}

\noindent We note that we have to evaluate two numerical integrals to calculate  $\partial_{\parallel}P_g   $. In order to find the pressure $P_g  (\parallel)$ we integrate equation (\ref{ode4}) using atmospheric pressure for some large value for $\parallel_{\infty}$ well outside the thin air gap as a boundary value. As a check of our analysis we now orientate the ($\parallel$,$\perp$)-coordinate system in such away that $\parallel=r$, to recover the lubrication equation in the conventional ($r$,$z$)-coordinate system. In that case we have $\theta=-\pi/2$, and we can write for $a(\parallel)$:

\begin{equation}\label{ode6}
a(\parallel=r)=\frac{\sin\theta}{r}=-\frac{1}{r}
\end{equation}

\noindent The integrating factor $I$ now becomes:

\begin{equation}\label{ode7}
I(\parallel=r)=\mathrm e^{-\int_{0}^{r}a(\tilde r)d\tilde r}=e^{\ln r}=r
\end{equation}

\noindent Substituting equation (\ref{ode7}) into equation (\ref{ode4}) and using the proposition $\parallel=r$ and setting $U_b=0$ and $U_d=0$, we can now write equation (\ref{ode4}) as:

\begin{equation}\label{ode8}
\partial_r P_g =-\frac{12\eta_g}{h^3}\left( \frac{1}{I(r)} \left( \int_{0}^{r} I(\tilde r)f(\tilde r)\,\dx\tilde r  \right)-\frac{h}{2}\left( U_{b}+U_{d} \right) \right)=\frac{12\eta_g}{h^3}\left( \frac{1}{r} \left( \int_{0}^{r} \tilde r \partial_{t}h\,\dx\tilde r  \right)\right) 
\end{equation}

\noindent We inspect that this equation (\ref{ode8}) is the equation for the radial pressure gradient for viscous lubrication theory in the conventional ($r$,$z$)-coordinate system~\citep{Bouwhuis_2013}, which gives a consistency check for our analysis. This was also numerically verified.  

\section{Results}
\label{sec:results}

In this section simulation results will be discussed, starting with section~\ref{sec:drop_pool} in which the drop impact onto a pool will be treated. The interface deformations and pressure development in the viscous air layer will be quantified. In section~\ref{sec:sphere} we will focus on rigid sphere impact onto a pool. For both impact scenarios we will quantify the size of the air bubble that is entrapped and directly compare with various experimental results~\citep{Tran_2013,Marston_2011,Bouwhuis_2012}. In section~\ref{sec:symm} we will compare the dynamics of both impact scenarios and identify symmetrical behavior. 

\subsection{Drop impact onto a pool}
\label{sec:drop_pool}
\begin{figure}
	\centerline{\includegraphics[width=1\textwidth]{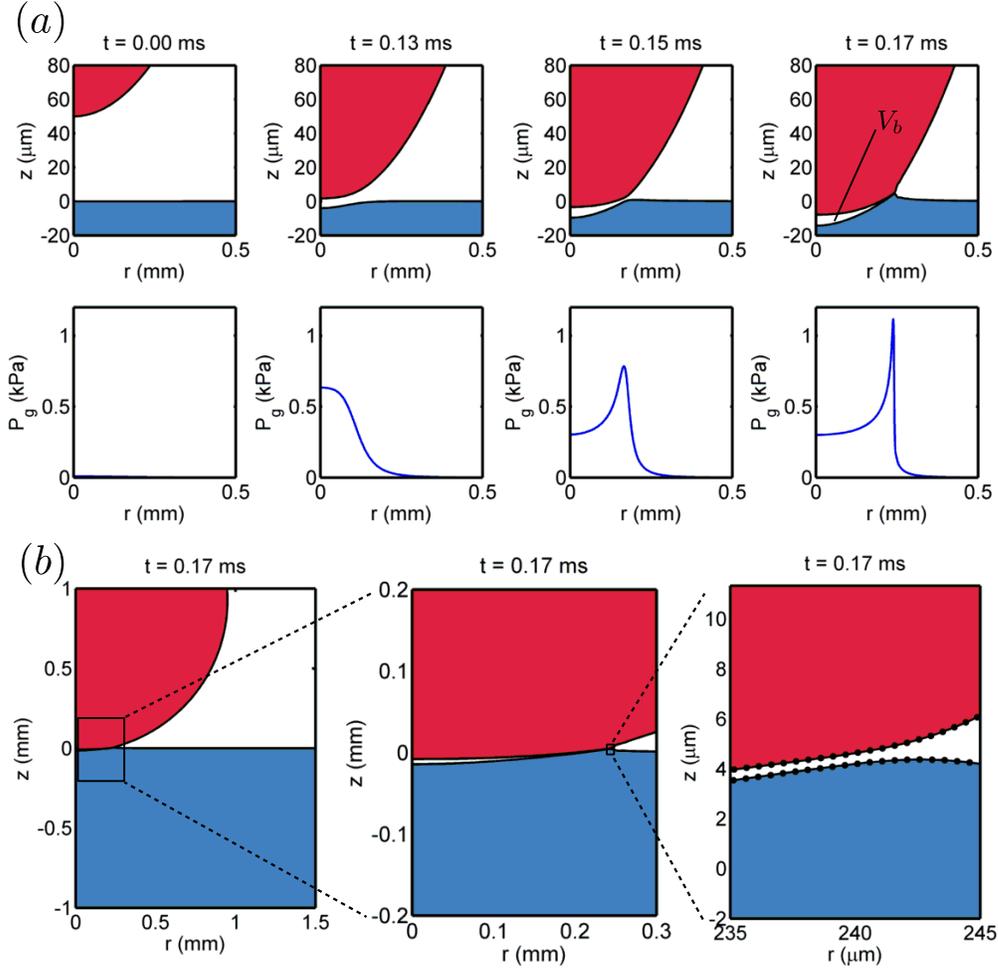}}
		\caption{(a) Drop impact onto a liquid pool. Note the different length scales for the $r$-axis and $z$-axis in the shape plots. The impact speed is $U=0.42\;\mathrm{m/s} $ and the drop radius is $R=0.95\;\mathrm{mm}$. The density and surface tension of the liquid are respectively $\rho=916\;\mathrm{kg/m^3} $ and $\gamma=0.020\;\mathrm{N/m} $. These impact parameters correspond to $\St= 2.0 \times 10^4 $ and $\We=7.7$.  The simulation starts at time $t=0\;\mathrm{ms} $ at a separation of $h_{r=0}=50\mum$. Due to the approach of the sphere, the excess air pressure $P_g$ will increase and acts on both the drop and the liquid pool ($t=0.13\;\mathrm{ms}$). At the final stage ($t=0.17\;\mathrm{ms}$) the minimum separation of the interfaces reaches $0.4\mum$ and the simulation is stopped. The bubble volume $V_b$ can thus be determined. (b) Part of the simulation domain with detailed snapshots of the air film at $t=0.17\;\mathrm{ms}$.  In the third snapshot the actual node distribution around the smallest separation point can be inspected. This is the most refined distribution of computational nodes that is used. For the region outside the gap a coarser node distribution is sufficient.}
	\label{fig:pool_evo}
\end{figure}

Figure~\ref{fig:pool_evo}a displays a typical result for drop impact onto a pool. The results are expressed in dimensional form to match the experimental conditions of the work of~\cite{Tran_2013}, to which the numerical results in this work will be compared. In the first frame corresponding to $t=0\;\mathrm{ms}$ the initial condition of the simulation at the impact zone is shown. An initial separation of $h_0=50\mum$ is used. Convergence tests regarding the initial release height have been conducted, and an initial separation of $h_0=50\mum$ was found to be appropriate for the lubrication pressure to be still negligible at this distance for the parameter range which is of interest in this study. At $t=0.12\;\mathrm{ms}$  it can be seen that the pool and the drop experience the increased air pressure and thus the interfaces deform. In the lower panel of this frame, the increase in pressure is indeed visible. At $t=0.15\;\mathrm{ms}$ the drop is getting closer to the pool, and the interfaces have been further deformed. It can also be noted that the pressure maximum corresponds to a location where the separation between the drop and the pool is smallest. The location of smallest separation is now not located in the center at $r=0$ anymore. This behavior is typical for impact events involving a free surface and has been experimentally observed for e.g. drop impact onto a pool~\citep{Thoroddsen_2012,Tran_2013}, drop impact onto a solid surface~\citep{Veen_2012,Ruiter_2012}, sphere impact onto a pool ~\citep{Marston_2011} and bubble impact onto a wall in a liquid tank~\citep{Hendrix_2012}. In the final frame $t=0.17\;\mathrm{ms} $ we observe that the two interfaces are very close together having a minimum separation of $0.4\mum$.

\begin{figure}
	\centerline{\includegraphics[width=1\textwidth]{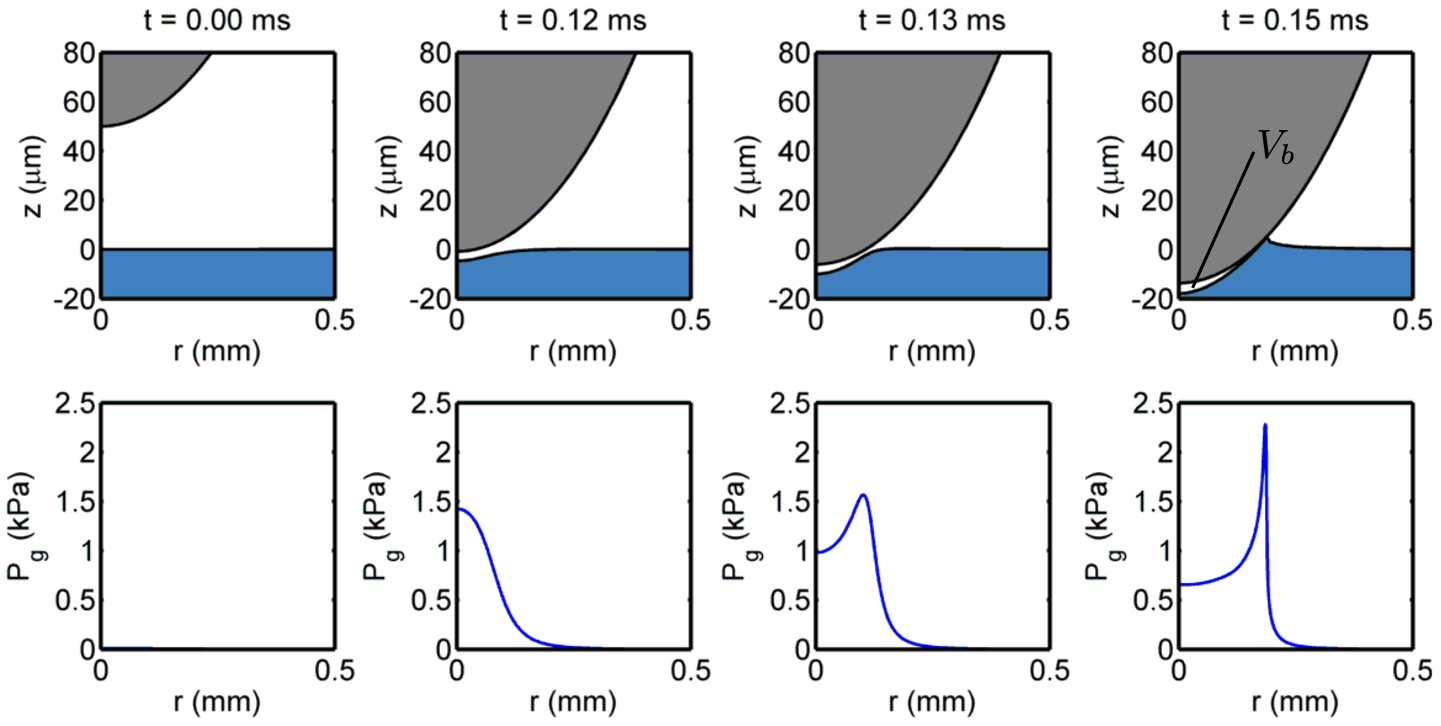}}
		\caption{Rigid sphere impact onto a pool. The impact speed is $U=0.42\;\mathrm{m/s} $ and the radius is $R=0.95\;\mathrm{mm}$. The density and surface tension of the fluid are respectively $\rho=916\;\mathrm{kg/m^3} $ and $\gamma=0.020\;\mathrm{N/m} $. These impact parameters correspond to  $\St= 2.0 \times 10^4 $ and $\We=7.7$. }
	\label{fig:sphere_evo}
\end{figure}

We note that the interfaces up to the final stage of impact are very well resolved, see figure~\ref{fig:pool_evo}b in which the final frame at $t=0.17\;\mathrm{ms}$ is shown on various scales while keeping both axes the same length scale. In the first frame in this figure a macroscopic view of the simulation domain is shown. In the second frame the impact zone is selected and magnified. The slender geometry of the microscopic air film can be noted. In the third frame the region of closest separation is magnified. Indeed, the interfaces are very close together, the minimum separation is $0.4\mum$. The computational nodes used for discretization of the surface are also shown in this final frame. An adaptive grid on the fluid surface allows for local refinement at the region of closest separation which results in the total number of nodes to be only of order 100, while capturing both the microscopic dynamics at the impact zone and the large scale motion of the millimeter sized drop. Note that the slanted orientation of the free surfaces in figure~\ref{fig:pool_evo}b justifies the need of using a description in terms of the ($\parallel$,$\perp$)-  rather than the ($r$,$z$)-coordinate system. 

We further note from the final frame in figure~\ref{fig:pool_evo}a that a microscopic air film finds itself trapped between the drop and the pool. It is this entrapped air that constitutes the air bubble that is dragged into the liquid when the air film ruptures at the thinnest point and breaks the axisymmetry of the problem. In this work we do not attempt to simulate the complex rupture process of the air film itself, which is ultimately determined by surface chemistry, see for example~\cite{Saylor_2012} who point out the importance of cleanliness of the fluid surface with respect to film rupture. Instead, we focus on the dynamics up to the rupture point, which is taken to happen at a rupture height of $0.4\mum$.  At this point, the volume of the entrained air has converged and can thus be determined, see the final frame of figure~\ref{fig:pool_evo}a. This procedure is in line with previous research~\citep{Bouwhuis_2012}, where experimentally the volume of the air pocket just before rupture was indeed found identical to the volume of the entrapped bubble.  

\subsection{Rigid sphere impact onto a pool}
\label{sec:sphere}

The impact of a sphere onto a pool prior to coalescence is similar to the case of a drop impacting on a pool, except that in the case of an impacting sphere the deformability of the impacting object is zero. This scenario has been simulated by letting an undeformable sphere approach the pool. The same equations are solved as described in Section~\ref{sec:numerical}, except no BIM is needed for the impacting sphere since the interface of the sphere is fixed. The result is depicted in figure~\ref{fig:sphere_evo}. Just as in the case of drop impact onto a pool, a microscopic air bubble is entrapped. As can be inspected, the air bubble has a similar shape, but its size is smaller than in case a drop impacts onto the pool, as can be inferred from a comparison to figure~\ref{fig:pool_evo}a. 

\begin{figure}
	\centering
	\includegraphics[width=1\textwidth]{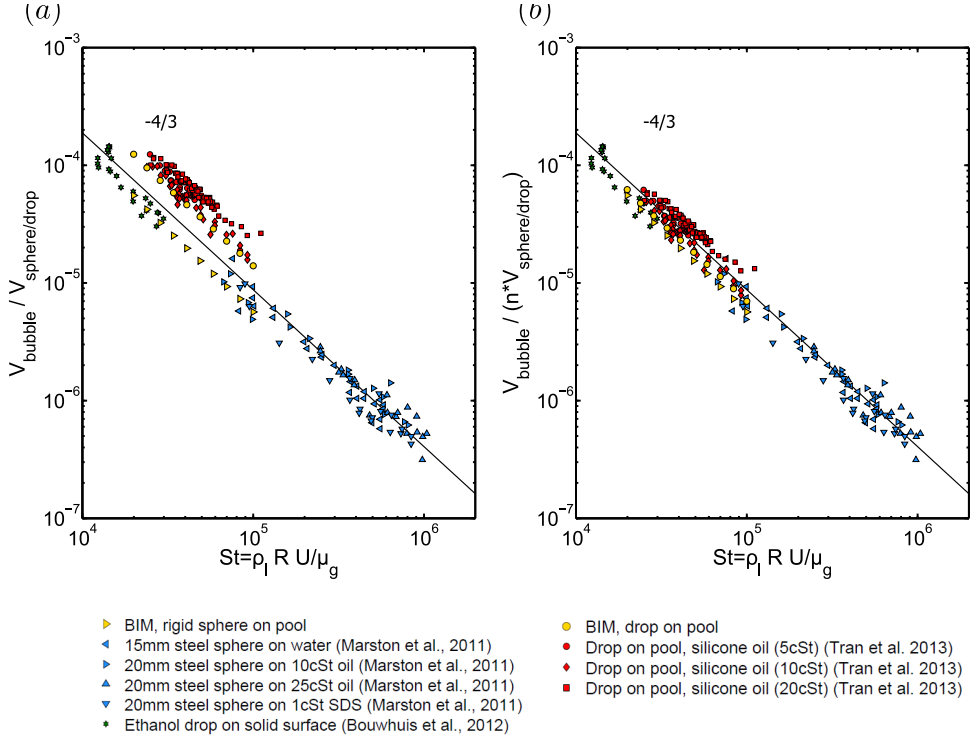}
		\caption{Figure adapted from al.~\cite{Tran_2013}. BIM results are superimposed in yellow symbols. (a) Various experimental data for the normalized bubble volume $V_b/V_{sphere/drop}$ are shown. Excellent quantitative agreement was found with numerical results. (b) The data, both numerical and experimental, was found to collapse on one single curve by normalizing $V_b$ as $V_b/(nV_{sphere/drop})$ with $n$ the number of free interfaces involved during impact. This number is 2 instead of 1 in case of drop impact onto a liquid pool.}  
		
	\label{fig:tuan}
\end{figure}

The size of the air bubble can be quantified from the numerical simulation and is compared for both drop impact and sphere impact onto a pool with various experimental results in figure~\ref{fig:tuan}a. We see that the numerical results of both drop and sphere impact onto a pool are in quantitative agreement with experimental work. We note that the numerical results show that the air bubble volume is indeed larger when a drop instead of a sphere impacts onto a pool for all $\St$, which is supported by experiments of~\cite{Tran_2013}. Furthermore, we observe that numerical results are in agreement with the scaling law presented in equation (\ref{vb2}), $V_b/V_{drop}\sim St^{-4/3}$. As experiments have shown, in this regime, viscosity of the liquid is not important for the final bubble volume that is entrapped, see~\cite{Marston_2011,Tran_2013}, which is again confirmed by the current modeling technique which captures the essential physics which determine the air bubble volume: a potential flow calculation that does not involve liquid viscosity coupled to viscous lubrication theory for the intervening airlayer.




%

\subsection{Deformations of interfaces: symmetrical behavior}	
\label{sec:symm}
\begin{figure}
	\centering{\includegraphics[width=1\textwidth]{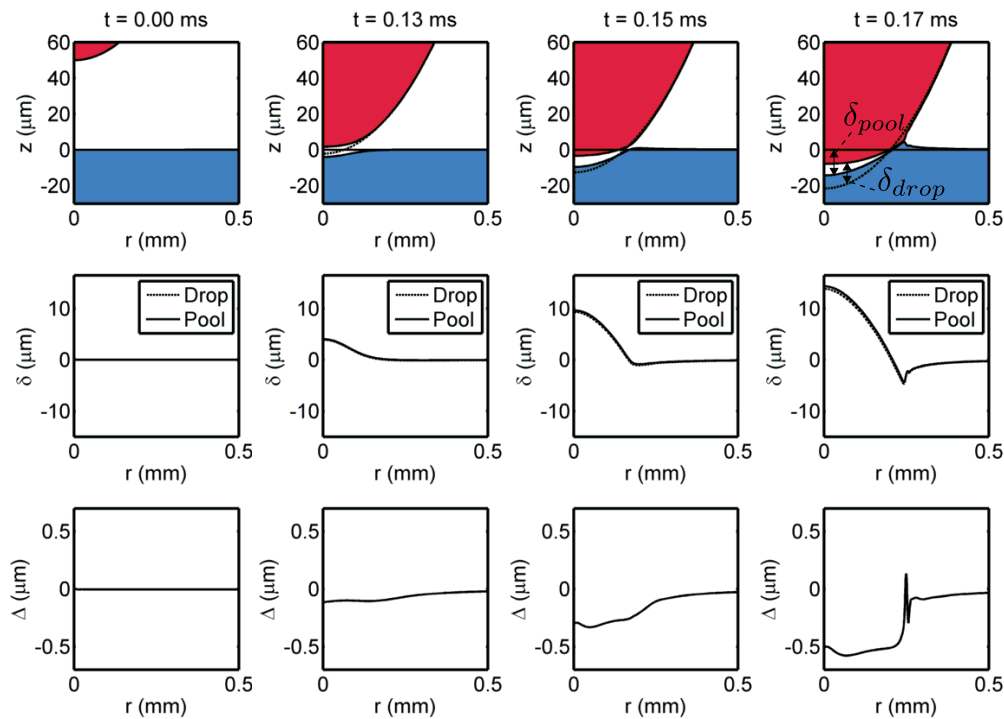}}
		\caption{Drop impact onto a pool with a corresponding plot of the relative deformation $\delta$ of both the pool and the drop. In the final frame of the upper panel the definition of $\delta$ is shown. We observe excellent overlap between the relative deformations, which is emphasized in the lower panel where $\Delta=\delta_{drop}-\delta_{pool}$, the difference between the two relative deformations, is shown.  The same impact conditions as for the case described in figure~\ref{fig:pool_evo} are used: The impact speed is $U=0.42\;\mathrm{m/s}$ and the radius is $R=0.95\;\mathrm{mm}$. The density and surface tension of the liquid are respectively $\rho=916\;\mathrm{kg/m^3}$ and $\gamma=0.020\;\mathrm{N/m}$, which corresponds to  $\St= 2.0 \times 10^4 $ and $\We=7.7$. }
	\label{fig:delta}
\end{figure}

\begin{figure}
	\centering	
	\includegraphics[width=1.0\textwidth]{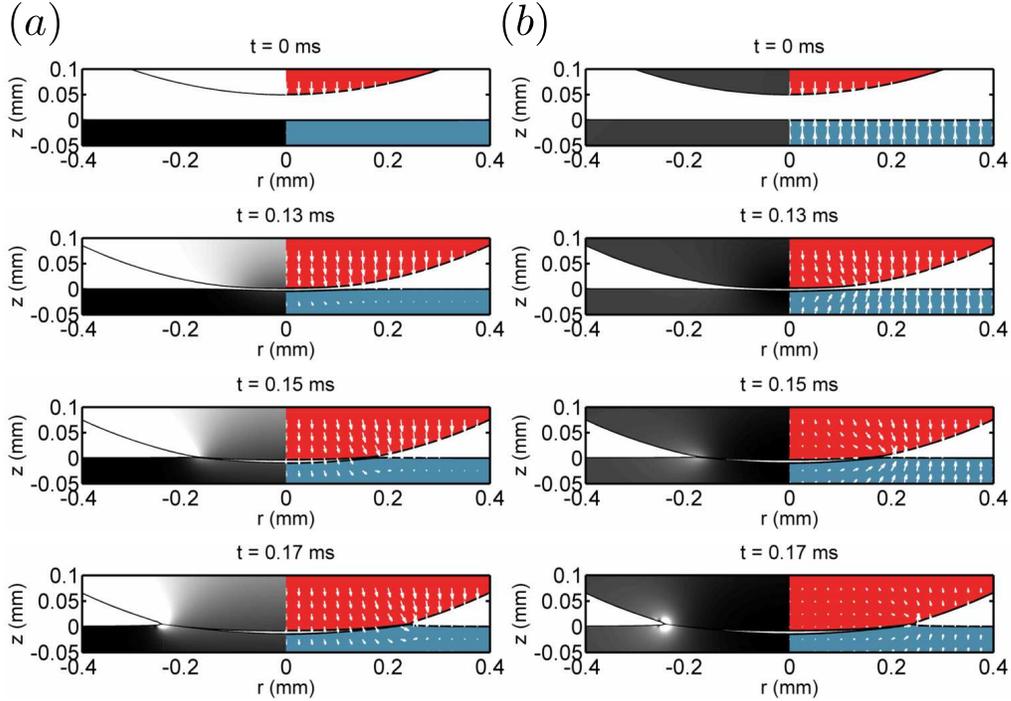}
		\caption{Kinetic energy monitoring during drop impact onto a pool, with impact parameters as described in figure~\ref{fig:pool_evo}. (a) In the left half, the kinetic energy $K$ is color coded: black is zero kinetic energy, white is maximum kinetic energy in the system which is $K=\frac{1}{2}\rho U^2$. (b) The kinetic energy is recalculated in a moving reference frame moving upwards at $\frac{1}{2}U$. This results in a frame of reference in which both the pool and drop move with a speed of $\frac{1}{2}U$ toward each other. Again the left half of the figure shows the kinetic energy. We observe a symmetric behavior which supports the hypothesis that the pool and drop react in a symmetric way to the local pressure increase.}
	\label{fig:energy}
\end{figure}

We will now further investigate the fact that the bubble volume for drop impact onto a pool is larger compared to the case where we deal with only one deformable interface during impact as is the case with rigid sphere impact onto a pool. In figure~\ref{fig:delta} a closer inspection of  the drop impact onto a pool is depicted. In this figure we track the relative deformation of both the pool and the drop, denoted by $\delta_{drop}$ and $\delta_{pool}$ respectively. Here $\delta_{drop}$ is defined as the deformation of the drop relative to an undeformed sphere impacting with constant speed $U$ and $\delta_{pool}$ is defined as the deformation of the pool relative to horizon $z=0$. Interestingly, we note that both interface deformations behave identically. One may expect that two deformable interfaces which react similar to an external pressure, deform in an identical way. Note, however, that the upper domain (drop) and the lower domain (pool) do not have the same unperturbed geometry, owing to the radius of curvature of the drop. Since both media respond identically to the pressure pulse, the weak curvature with respect to the width of the localized pressure has a negligible influence: on the scale of the pressure pulse, both domains are essentially flat. We therefore expect to recover a symmetric response in the upper and lower domains. To illustrate this further, we compute the kinetic energy and the velocity inside the drop and the pool using a technique described in~\citet{Sun_2014} to evaluate quantities close to the interface which need special attention as the singular behavior of the Green's function in the Boundary Integral equation becomes apparent for these points. Figure~\ref{fig:energy}a shows the result in the frame of the pool. To highlight the symmetry, we also evaluate these quantities in a frame moving at a speed $U/2$ in an upward direction, which results in a frame of reference in which both the drop and pool move with a speed $U/2$ towards each other. Indeed, the velocity fields and kinetic energies are now identically distributed, see figure~\ref{fig:energy}b.

This implies that there will be a bigger entrapped air bubble as compared to the case where only one of the interfaces is able to deform. To quantify this hypothesis we compare the bubble sizes of drop and sphere impact onto a pool and find a factor 2 difference, see figure~\ref{fig:tuan}b. Here half the air bubble volume of drop impact onto a pool was found to collapse onto the experimental and numerical results incorporating only one deformable interface, i.e. the sphere impact onto a pool but also drop impact onto a solid.~\cite{Tran_2013} took another approach to collapse the data of bubble volumes of drop impact onto a pool by correcting the corresponding impact $\St$ number by a factor 2, which also collapses the data. In this present work it is shown that an approach based on considering the number of deformable interfaces (either 1 or 2) can also serve to obtain a unifying view on the air bubble entrapment.

\section{Conclusion} 
\label{sec:conclusion}
In this work air entrapment during liquid drop and rigid sphere impact onto a deep liquid pool has been numerically investigated using a Boundary Integral Method (BIM) for potential flow for the liquid phase coupled to the viscous lubrication approximation for the subphase air which is squeezed out during impact. Excellent agreement with experimental work was found when comparing the amount of air that is entrained during impact. When considering drop impact onto a pool both liquid interfaces were found to deform identically relative to their undeformed shape. This leads to an explanation as to why bubble volumes in case of drop impact onto a pool were found to be twice the size of those that are found, both experimentally and numerically, in impacts events involving only one deformable interface, that is, rigid sphere impact onto a pool and drop impact onto a solid. In this study compressibility effects of the air have been neglected. It can be expected that at higher impact velocity compressibility of the intervening air will be important, see for example~\citet{Hicks_2011}. In addition, the current modelling technique is limited to an axisymmetric 2D framework. To account for 3D impact problems, for which experimental data starts to emerge~\citep{Veen_2014}, the modelling technique needs to be extended to 3D. With a 3D model also oblique collisions can be investigated. 

\appendix
\section{Continuity in curvilinear coordinates}
\label{sec:app}

To derive equation~\ref{continuity_tn} in a ($\parallel$,$\perp$)-coordinate system that moves along with the drop surface, see figure~\ref{fig:tn}, we start from the continuity equation in axisymmetric ($r$,$z$) coordinates:

\begin{equation}
\label{eq:cont_rz}
\frac{u_r}{r}+\frac{\partial u_r}{\partial r}+\frac{\partial u_z}{\partial z}=0.
\end{equation}

\noindent We now want to write the last two terms of the LHS of equation~\ref{eq:cont_rz} in terms of the ($\parallel$,$\perp$)-coordinate system, that is:

\begin{equation}
\label{cont_ns0}
\frac{u_r}{r}+\frac{\partial u_r}{\partial r}+\frac{\partial u_z}{\partial z}= \frac{u_r}{r}+\left(  \frac{\partial u_r}{\partial \perp}\frac{\partial \perp}{\partial r}+\frac{\partial u_r}{\partial \parallel}\frac{\partial \parallel}{\partial r}\right ) +  \left(  \frac{\partial u_z}{\partial \perp}\frac{\partial \perp}{\partial z}+\frac{\partial u_z}{\partial \parallel}\frac{\partial \parallel}{\partial z}\right )
\end{equation}
	
\noindent The two coordinate systems are related as follows (see also figure~\ref{fig:tn}):

\begin{align}
\label{s_n}
\parallel=-r\sin\theta  + z\cos\theta 
\\
\perp=r\cos\theta + z\sin\theta 
\end{align}

\noindent Using the relation above we can write equation~\ref{cont_ns0} as:
		
\begin{equation}
\label{cont_ns1}
\frac{u_r}{r}+\frac{\partial u_r}{\partial r}+\frac{\partial u_z}{\partial z}= \frac{u_r}{r}+\left(  \frac{\partial u_r}{\partial \perp}\cos\theta -\frac{\partial u_r}{\partial \parallel}\sin\theta \right ) +  \left(  \frac{\partial u_z}{\partial \perp}\sin\theta+\frac{\partial u_z}{\partial \parallel}\cos\theta\right )
\end{equation}		

\noindent We now have to express $u_r$ and $u_z$ as function of ($\parallel$,$\perp$), that is:

\begin{align}
\label{ur_uz}
u_r(\perp,\parallel)=u_{\perp}(\perp,\parallel)\cos\theta  - u_{\parallel}(\perp,\parallel)\sin\theta  
\\
u_z(\perp,\parallel)=u_{\perp}(\perp,\parallel)\sin\theta  + u_{\parallel}(\perp,\parallel) \cos\theta 
\end{align}

\noindent Substituting the above expressions for $u_r$ and $u_z$ into equation~\ref{cont_ns1} and simplifying we find:

\begin{equation}
\label{cont_ns4}
\frac{u_r}{r}+\frac{\partial u_r}{\partial r}+\frac{\partial u_z}{\partial z}= \frac{u_r}{r}+\frac{\partial u_{\perp}}{\partial \perp}+\frac{\partial u_{\parallel}}{\partial \parallel}
\end{equation}

%

\section*{Acknowledgments}
We gratefully acknowledge Jeremy Marston and Tuan Tran for providing their original experimental data set. This work was supported by STW and NWO
through a VIDI Grant No. 11304.

\bibliographystyle{jfm}

\end{document}